\documentclass[range]{ar2e}
\usepackage{epsfig}
\begin{document}

\renewcommand\baselinestretch{1}


\input epsf.tex    
\newcommand{\ra}{\rightarrow}
\newcommand{\rt}{\rightarrow}
\newcommand{\etal}{{\em et al.}}

\input psfig.sty

\jname{..}
\jyear{2000}
\jvol{}
\ARinfo{1056-8700/97/0610-00}

\title{The Exotic $XYZ$ Charmonium-like Mesons}

\markboth{Godfrey \& Olsen}{The Exotic $XYZ$ Charmonium-like Mesons}

\author{Stephen Godfrey
\affiliation{Ottawa-Carleton Institute for Physics \\
Department of Physics, Carleton University, Ottawa,  Canada K1S 5B6}
Stephen L. Olsen
\affiliation{Institute of High Energy Physics, Beijing, 100049 China, 
and\\
Department of Physics \& Astronomy\\
University of Hawaii at Manoa, Honolulu, Hawaii, 96822, U.S.A.
}}

\begin{abstract}
Charmonium, the spectroscopy of $c\bar{c}$ mesons, has recently enjoyed a 
renaissance with the discovery of several missing states and a number 
of unexpected charmonium-like resonances.  
The discovery of these new states has been made possible by the 
extremely large data samples made available by the $B$-factories at the 
Stanford Linear Accelerator Center and at KEK in Japan, and at the CESR 
$e^+e^-$ collider at Cornell. 
Conventional $c\bar{c}$ states are well described by quark potential 
models.  However, many of these newly discovered charmonium-like mesons do 
not seem to fit into the conventional $c\bar{c}$ spectrum.  There is 
growing evidence that at least some of these new states are {\it exotic},
{\it i.e.} new forms of hadronic matter such as mesonic-molecules, 
tetraquarks, and/or hybrid mesons.  
In this review we describe expectations for the properties of conventional
charmonium states and the predictions for molecules, tetraquarks and 
hybrids and the various processes that can be used to produce them.  We 
examine the evidence for the new candidate exotic mesons, possible 
explanations, and experimental measurements that might shed further 
light on the nature these states. 

\end{abstract}

\maketitle

\baselineskip 0.5 true cm
\renewcommand\baselinestretch{1}

\section{Introduction}

In 1964, faced with a large proliferation of
strongly interacting subatomic particles,
Murray Gell-Mann~\cite{Gell-Mann} and, independently, 
George Zweig~\cite{Zweig} 
hypothesized the existence of  three fractionally
charged constituent fermions called "quarks," the
charge$=+2/3e$  up-quark ($u$) and the charge$=-1/3e$ down- ($d$) 
and strange- ($s$) quarks and their antiparticles 
($e=+1.6\times 10^{-19}$ Coulombs).
In the Gell-Mann Zweig scheme, 
mesons are formed from quark and antiquark ($q\bar{q}$) pairs 
\footnote{Meson: a bound state of a quark and antiquark $(q\bar{q})$}
and
baryons from three-quark triplets ($qqq$)
\footnote{Baryon: a bound state of three quarks $(qqq)$}.  
This picture was remarkably successful; it 
accounted for  all of the known hadrons at that time and 
predicted the existence of additional hadrons that were
subsequently discovered.  

Now, over forty years later, the Gell-Mann Zweig idea, currently
known as the ``Constituent Quark Model'' (CQM),
\footnote{CQM: constituent quark model}
remains an effective
scheme for classifying all of the known hadrons, although the 
number of quarks has expanded to include the charge$=+2/3e$ 
charmed- ($c$) and top- ($t$) quarks and the charge$= -1/3e$ bottom- ($b$) 
quark.

Our current understanding is that 
the forces that bind quarks
into hadrons are described by the non-abelian field theory
called Quantum Chromodynamics (QCD)
\footnote{QCD: Quantum Chromodynamics, the theory of the strong interactions}.  
At distance scales that correspond to the separations between quarks
inside hadrons, QCD is a strongly coupled theory and perturbation 
theory is of limited applicability.  It is expected that
ultimately, numerical lattice QCD computations \cite{Bali:2000gf} will
generate predictions for QCD observables such as masses, 
transitions, decays, etc. However, progress, while steady, is slow,   
and it will be some time before the predictions are able
to make precise reliable predictions for excited charmonium states.
\footnote{Lattice QCD: A numerical approach to calculate hadronic 
properties.  In the lattice approach
space-time is discretized and observables are typically calculated using Monte Carlo techniques 
used to calculate expectation values of various operators by integrating
over the quark gluon configurations.}
To date, models that incorporate general features 
of the  QCD theory have proven to be most useful for describing
the spectra and properties of hadrons.
A prediction of these QCD-motivated models, supported by lattice QCD,
is the existence of hadrons with more complex substructures
than the simple $q\bar{q}$ mesons and the $qqq$ baryons.  
However, in spite of considerable experimental effort,
no unambiguous evidence for hadrons with a non-CQM-like
structure has been found,  at least not until recently,
when studies of the spectrum of charmonium mesons, {\it i.e.}
mesons formed from a $c\bar{c}$ pair, have uncovered a
number of meson candidates that do not seem to conform
to CQM expectations.  The status of these candidate non-$q\bar{q}$
particles, the 
so-called $XYZ$ mesons, is the subject of this review, which
is organized as follows: Section 2 provides a brief
summary of theoretical expectations for charmonium
mesons and the more complex structures that are expected
in the context of QCD; Section 3 provides some 
experimental background of the recent observations;
Section 4 
forms the bulk of the review, describing the evidence for and measured properties of the
$XYZ$ states,  why they defy any CQM assignment, and theoretical speculation about
their nature and how to test these hypotheses.
Summary points and future issues are given in Section 5.  
Recent related reviews on charmonium 
spectroscopy and the $XYZ$ states are given in
Refs.~\cite{Eichten:2007qx,Swanson:2006st,Godfrey:2006pd,Zhu:2007wz}.

\section{Theoretical Background}

QCD-motivated potential models successfully described the $J/\psi$ and $\psi'$
as 
$c\bar{c}$ states soon after they were discovered 
more than thirty years ago.  These models have stood up quite well over 
the ensuing years, during which time other low-lying $c\bar{c}$ 
states were discovered and found
to have properties that agree reasonably well with the models' predictions.  
Early pioneering papers also predicted the existence of mesons
that are more complicated than  conventional $q\bar{q}$ states 
such as multiquark states~\cite{Voloshin:1976ap,DeRujula:1976qd}
\footnote{Multiquark state: a state that has more quark or antiquark content
than conventional $q\bar{q}$ mesons and $qqq$ baryons}
and hybrid mesons, states with an excited gluonic degree of freedom 
\cite{Buchmuller:1979gy}
\footnote{Hybrid meson: a meson with an excited gluonic degree of freedom}.  
While corresponding
{\it exotic} states are predicted to exist in the light meson spectrum,
there they are 
difficult to disentangle from the dense background of conventional 
states~\cite{Godfrey:1998pd}.  
The charmonium spectrum provides a cleaner environment where
one might hope that non-conventional states containing $c\bar{c}$ 
pairs would be easier to identify.  In this section we give a brief 
overview of the properties of conventional charmonium 
states and the non-conventional multiquark and hybrid states.  
In addition, we mention threshold
effects that could masquerade as resonances.

\subsection{Charmonium states and properties}

\begin{figure}[htb!]
\centerline{\epsfysize 3.0 truein
\epsfbox{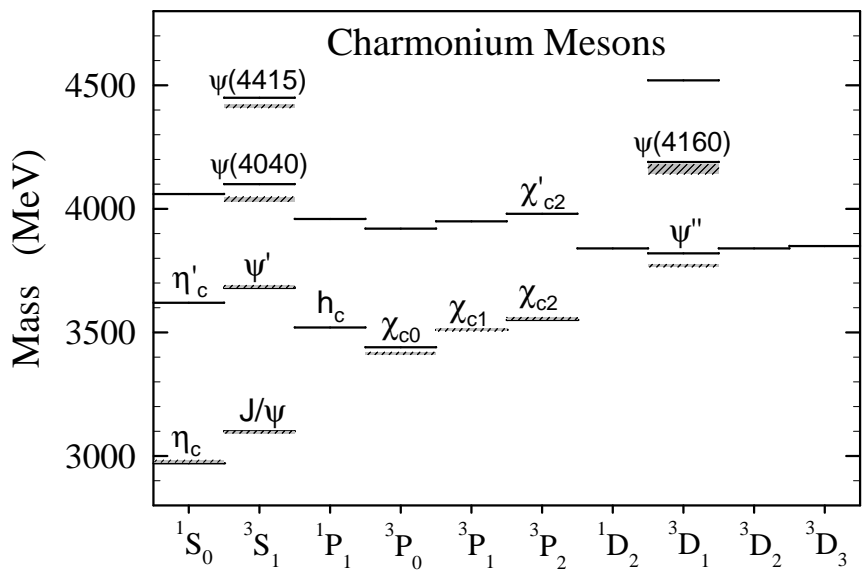 }}
\caption{The charmonium level diagram.  The commonly
used names for the mesons associated with assigned
states are indicated.
}
\label{fig:ccbar-spectrum}
\end{figure}

In QCD-motivated quark potential models, 
quarkonium states are described as a quark-antiquark pair 
bound by an inter-quark force with a short-distance behavior 
dominated by single-gluon-exchange and, thus, approximately Coulombic,
plus a linearly increasing confining potential that dominates at large separations.
Typically, the energy levels are found by solving a non-relativistic 
Schr\"{o}dinger equation, although there are more sophisticated 
calculations that take into account 
relativistic 
corrections and other effects.  These give energy levels 
that are characterized by the radial quantum number $n$ and the
relative orbital angular momentum  between the quark and antiquark, $L$.
The current status of this approach is shown in Fig.~\ref{fig:ccbar-spectrum},
where the charmonium levels are shown.  For those levels that have been assigned,
the commonly used name of its associated meson is indicated.

The orbital levels are labeled by $S$,
$P$, $D$, \ldots corresponding to $L=0$, 1, 2 \ldots.  The quark and antiquark spins
couple to give the total spin $S=0$ (spin-singlet) or $S=1$ (spin-triplet). 
$S$ and $L$ couple to give the total angular momentum of the state, $J$.  
The parity of a quark-antiquark
state with orbital angular momentum $L$ is $P=(-1)^{L+1}$ and the charge conjugation eigenvalue 
is given by $C=(-1)^{L+S}$.  Quarkonium states are generally denoted by $^{2S+1}L_J$ with quantum 
numbers $J^{PC}$.  Thus, the $L=0$ states are  $^1S_0$ and $^3S_1$ with $J^{PC}= 0^{-+}$ and 
$1^{--}$, respectively; the $L=1$ states are $^1P_1$ and $^3P_{0,1,2}$ 
with  $J^{PC}= 1^{+-}$,  
$0^{++}$,  $1^{++}$, and $2^{++}$; the $L=2$ states are $^2D_2$ and 
$^3D_{1,2,3}$ with  $J^{PC}= 2^{-+}$,  $1^{--}$,  $2^{--}$, and $3^{--}$, etc.

In addition to the spin-independent potential, there are spin-dependent interactions that 
produce corrections of order $(v/c)^2$. These are found by assuming a specific 
Lorentz structure for the 
quark-antiquark interactions.  Typically the short distance one-gluon-exchange is taken
to be a Lorentz vector interaction and the confinement piece is assumed to be a
Lorentz scalar. This
gives rise to splittings within multiplets.  For example, the $J/\psi(1^3S_1)-\eta_c(1^1S_0)$ 
splitting
is attributed to a short distance $\vec{S}_Q\cdot\vec{S}_{\bar{Q}}$ contact 
interaction arising from the one-gluon-exchange, 
while the splittings of the $P$-wave 
$\chi_c(1^3P_{J=0,1,2})$  and higher $L$ 
states are due to spin-orbit and tensor spin-spin interactions arising from one-gluon-exchange 
and a relativistic spin-orbit Thomas precession term. 
The recent measurement of the $h_c$ 
mass by CLEO~\cite{CLEO_hc} is an important validation of this picture.

An important approach to understanding the charmonium spectrum is lattice 
QCD \cite{Bali:2000gf}.  
There has been considerable recent progress in calculating the masses of
excited $c\bar{c}$ states and radiative transitions \cite{Dudek:2007nj},
 although  the results are still not at the point that they can make
precision predictions for excited states. However, lattice QCD calculations of the static 
energy between a heavy quark-antiquark pair are in 
good agreement with phenomenological potentials
\cite{Bali:2000gf}
and lattice calculations of the spin-dependent potentials also support the phenomenological contact,
tensor, and spin-orbit potentials \cite{Koma:2005nq}. 

All charmonium states below the $D\bar{D}$ 
``open-charm'' mass threshold\footnote{The $D^0$ 
and $D^+$ mesons are the spin-singlet $S$-wave ($1^1S_0$) $c\bar{u}$ 
and $c\bar{d}$ quark state, respectively.} 
have been observed.  The $D$-wave $c\bar{c}$ states 
lie just above this threshold and, thus far, only the $1^3D_1$ state, which 
is identified as the $\psi^{\prime\prime}$ (or $\psi(3770)$), has
been observed.  The remaining $n=1$ $D$-wave charmonium states,  
the spin triplet $^3D_2$ and $^3D_3$ states and the spin singlet $^1D_2$ state,
are all expected to have masses near 3800~MeV/$c^2$ and to be narrow
\cite{Barnes:2005pb,Barnes:2003vb,Eichten:2004uh,Eichten:2005ga} 
so that it should be possible to observe them \cite{Eichten:2002qv}.  
The $n=2$ $P$ states are the next 
highest multiplet, with masses predicted to lie in the range of 
$3800-3980$~MeV/$c^2$ and with total widths of 42, 165, 30, and 87~MeV/$c^2$ 
for the $2^3P_2$, $2^3P_1$, $2^3P_0$, and $2^1P_1$,
respectively.   We also mention the $1^3F_4$ since it is relatively 
narrow for a state so massive, with a predicted mass and width of 4021 and 
$\sim 8$~MeV respectively.  
The $\eta_c''$ ($3^1S_0$) mass and width are predicted to 
be $\sim 4050$~MeV/$c^2$ and $\sim 80$~MeV/$c^2$.

\subsubsection{Transitions and decays}

While the mass value is an important first element for the identification 
of a new state, more information is usually needed to distinguish between 
different possibilities. One therefore needs 
other measurements to form a detailed picture of its internal structure. 
Decay properties are an important aspect of this.

Electromagnetic transitions can potentially give information on
the quantum numbers of a parent state when it decays to a final state with 
established quantum numbers such as the $J/\psi$ or $\chi_{cJ}$;  studies 
of the angular correlations among the final state 
decay particles provide additional information.  
Quark model predictions also provide an important benchmark against which to test a 
conventional quarkonium interpretation versus an exotic one. 
The theory of 
electromagnetic transitions between quarkonium states is straightforward, and potential models provide
detailed predictions that can be compared to experimental measurements to test the internal
structure of a state.  
The leading-order amplitudes are those for electric
($E1$) and magnetic ($M1$) dipole transitions, with the $E1$ amplitude being the most relevant 
to our 
discussion.  The predictions for the $^3P_J \leftrightarrow ^3S_1$ transitions
are in good agreement with experimental data \cite{Eichten:2007qx}.  We 
can therefore expect that other electromagnetic transitions will also yield useful information, 
with the possible exceptions for those cases where
there are large dynamical cancellations in the matrix elements, for instance 
those that involve higher radial excitations that have nodes in their wavefunctions.
The status of 
electromagnetic transitions has been reviewed recently \cite{Eichten:2007qx} and detailed predictions 
for charmonium states are given in 
Refs.~\cite{Barnes:2005pb,Barnes:2003vb,Eichten:2004uh,Eichten:2005ga}.  

Charmonium states can also undergo hadronic transitions from one 
$c\bar{c}$ state to another via the emission of light hadrons. 
Examples 
of observed transitions include $\psi(2S)\to J/\psi \pi^+\pi^-$, $\psi(2S)\to J/\psi \eta$,
$\psi(2S)\to J/\psi \pi^0$, and $\psi(2S)\to h_c \pi^0$.  
The theoretical description of 
hadronic transitions uses a multipole expansion of the gluonic fields 
\cite{Gottfried:1977gp,G2,G3,G4,G5,Yan:1980uh} which resembles the usual
multipole expansion applied to electromagnetic transitions. 
Detailed predictions for hadronic transitions are given in 
Refs.~\cite{KY,Voloshin:2006ce,KYetc,KYetc2,KYetc3,KYetc4,KYetc5,KYetc6}.

Charmonium states above the $D\bar{D}$ 
and/or $D\bar{D^*}$ mass
threshold can decay to $D\bar{D^{(*)}}$ final 
states.\footnote{The $D^{*}$ mesons are the spin-triplet $1^3S_1$ partners
of the $D$ mesons.}
These decays
are well described by the $^3P_0$ model in which a light $q\bar{q}$ pair is created out of the 
vacuum with the quantum numbers of the vacuum, $0^{++}$ 
\cite{Micu:1968mk,LeYaouanc:1972ae,Barnes:2005pb,Barnes:2003vb}.  
The partial decay widths have been 
calculated for many mesons using this model and the overall qualitative agreement with experiment
is very good.  Thus,  the predictions can provide a useful means of identifying conventional 
charmonium states.  Recent calculations of decay properties of excited charmonium states are
given in Refs.~\cite{Barnes:2005pb,Barnes:2003vb,Eichten:2004uh,Eichten:2005ga}.

A final decay mechanism of charmonium states is 
via the annihilation of the $c\bar{c}$ into final states
consisting of gluons and light quark pairs, sometimes with a photon
\cite{KMRR,Petrelli:1997ge}.  
However, so far at least, these have not proven to be very important for understanding the new states.

\subsection{Multiquark states}

An early quark model prediction was the existence of multiquark states, specifically
bound meson-antimeson molecular states~\cite{Voloshin:1976ap,DeRujula:1976qd}.  
In the light quark sector the $f_0(980)$ and $a_0(980)$ are considered to
be strong candidates for $K\bar{K}$ molecules.  However, in general, it is challenging
to definitively identify a light multiquark state in 
an environment of many broad and often overlapping conventional states.
The charmonium spectrum is better defined so that new types of states can 
potentially be more easily
delineated from conventional charmonium states.  
The observation of the $X(3872)$, the first of the $XYZ$ particles to be seen, 
brought forward the hope
that one can definitively state that a multiquark state has been observed.  

\begin{figure}[htb!]
\centerline{\epsfysize 3.0 truein
\epsfbox{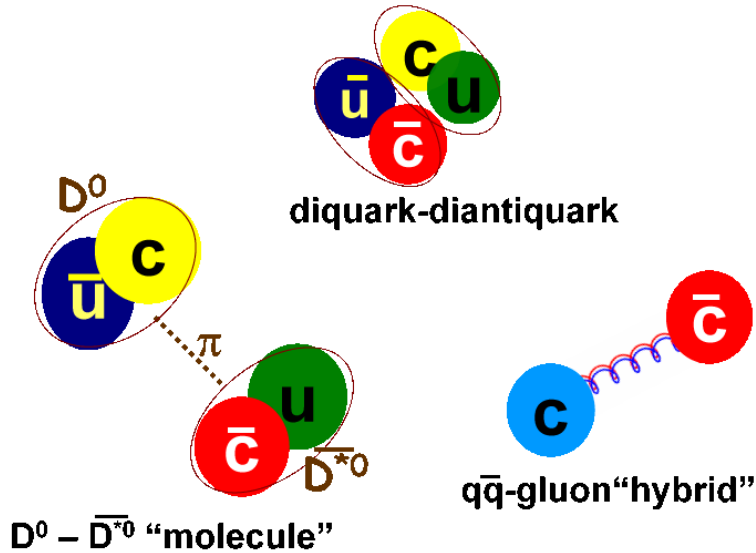 }}
\caption{Cartoon representations of molecular states,
diquark-diantiquark tetraquark mesons and 
quark-antiquark-gluon hybrids.}
\label{fig:cartoon_fig}
\end{figure}

Two generic types of multiquark states have been described in the literature. 
The first, a molecular state,
sometimes referred to as a deuson \cite{tornqvist_x3872}, is comprised of two
charmed mesons bound together to form a molecule.  
These states are by nature loosely bound.  
Molecular states bind through two mechanisms: quark/colour exchange 
interactions at short distances
and pion exchange at large distance \cite{Swanson:2006st,tornqvist_x3872,Ericson:1993wy}
(see Fig.~\ref{fig:cartoon_fig}) although pion exchange is expected to dominate \cite{Swanson:2006st}. 
Molecular states are generally not isospin eigenstates,
which gives rise to distinctive decay patterns.
Because the mesons inside the molecule are weakly bound, 
they tend to decay as if they are free. Details are reviewed by Swanson
in Ref.~\cite{Swanson:2006st}.

The second type is a tightly bound four-quark state, dubbed a tetraquark, 
that is
predicted to have properties that are distinct from those of a molecular state.
In the model of Maiani {\it et al} \cite{Maiani:2004vq},
the tetraquark is described as a diquark-diantiquark 
structure in which the quarks 
group into colour-triplet scalar and vector clusters 
and the interactions are dominated by a simple
spin-spin interaction (see Fig.~\ref{fig:cartoon_fig}).  
Here, strong decays are expected to proceed via rearrangement processes followed by dissociation 
that give rise, for example, to decays such as $X\to \rho J/\psi \to \pi\pi J/\psi$ or 
$X\to D\bar{D}^* \to D\bar{D}\gamma$.  

A prediction that distinguishes multiquark states containing a $c\bar{c}$ pair
from conventional charmonia is the possible existence of multiplets 
that include members with non-zero charge (e.g. $[cu\bar{c}\bar{d}]$), strangeness
(e.g. $[cd\bar{c}\bar{s}]$), or both (e.g. $[cu\bar{c}\bar{s}]$)~\cite{chiu:2006}.

\subsection{Charmonium hybrids}

Hybrid mesons are states with an excited gluonic degree of freedom (see
Fig.~\ref{fig:cartoon_fig}).  These
are described by many different models and calculational schemes~\cite{Barnes:1995hc}.
A compelling description, supported by lattice 
QCD~\cite{Morningstar:2000mf,Bali:2003jq}, views the quarks as
moving in adiabatic potentials produced by gluons in analogy to the atomic nuclei 
in molecules moving in the adiabatic potentials produced by electrons.
The lowest adiabatic surface leads to the 
conventional quarkonium spectrum while the excited adiabatic surfaces 
are found by putting the quarks into more complicated colour 
configurations.   
In the flux-tube model~\cite{Isgur:1984bm}, the lowest excited adiabatic 
surface corresponds to transverse excitations of the flux tube and 
leads to a doubly degenerate octet of the lowest mass hybrids with 
quantum numbers $J^{PC}=0^{+-}$, $0^{-+}$, $1^{+-}$, $1^{-+}$, 
$2^{+-}$, $2^{-+}$, $1^{++}$ and $1^{--}$. The $0^{+-}$, $1^{-+}$, $2^{+-}$
quantum numbers are not possible for a $c\bar{c}$ bound state
in the quark model and are referred 
to as {\it exotic} quantum numbers.  If observed, they would unambiguously 
signal the existence of an unconventional state. 
Lattice QCD and most models predict the 
lowest charmonium hybrid state to be roughly 4200~MeV/c$^2$ in mass
\cite{Isgur:1984bm,Barnes:1995hc,Lacock:1996ny}.

Charmonium hybrids can decay via electromagnetic transitions, hadronic 
transitions such as $\psi_g\to J/\psi +\pi \pi$, and to open-charm 
final states such as 
$\psi_g\to D^{(*,**)}\bar{D}^{(*,**)}$\footnote{$D^{**}$ denotes
mesons that are formed from $P$-wave $c\bar{q}$ ($q=u$ or $d$) pairs: 
$D_0^*(^3P_0)$, $D_2^*(^3P_2)$ and the $D_1$ and $D_1'$ are $^3P_1-^1P_1$ mixtures.}.
The partial widths have 
been calculated using many different models.  There are some
general properties that seem to be supported by most models and by recent 
lattice QCD calculations.  Nevertheless, since there are no experimental results 
against which to test these calculations, one should take their 
predictions with a grain of salt.  Two important decay modes are:
\begin{enumerate}
\item $\psi_g\to D^{(*,**)}\bar{D}^{(**,*)}$. 
Most calculations 
predict that the $\psi_g$ should decay to a $P$-wave plus an $S$-wave 
meson.  
In this case $D(L=0)+D^{**}(L=1)$ final states should 
dominate over decays to $D\bar{D}$ and the 
partial width to $D\bar{D}^*$ should be very small.
\item $\psi_g \to (c\bar{c})(gg)\to (c\bar{c})+(\pi\pi, \; \eta 
\ldots)$ These modes offer clean experimental signatures.  If the total width 
is small, they could have significant branching fractions.  One recent 
lattice QCD calculation finds that these types of decays are potentially quite 
large, ${\cal O}(10\hbox{ MeV}/c^2)$~\cite{McNeile:2002az}.
\end{enumerate}

\subsection{Threshold effects}

In addition to various types of resonances, 
thresholds can also give rise to structures in cross sections
and kinematic distributions. 
Possible thresholds include the $DD^*$,  $D^*D^*$, $DD_1$, $D^*D_1$  at 
$E_{cm}\sim$  38721, 4020, 4287, and 4430~MeV respectively.  
At threshold, the cross section is typically dominated by $S$-wave (L=0) 
scattering although in some cases higher waves can be important.
States in a relative $S$-wave with little relative momentum and who 
live long on the time scale of strong interactions will have enough time to 
exchange pions and interact \cite{Close:2008hv}.  
Binding is then possible via an attractive $\pi$ exchange which could occur via 
couplings such as $D\leftrightarrow D^*\pi^0$ and can lead to the molecular states 
discussed above.    
However, other strong interaction effects might also lead to a repulsive 
interaction that could result in a virtual state above threshold. Thus, 
passing 
through a kinematical threshold can lead to structure in the cross section that may or
may not indicate a resonance.  
In addition, if there are nearby $c\bar{c}$ states, they 
will interact with the threshold resulting in mass shifts of 
both the $c\bar{c}$ resonance and the threshold-related enhancement.  
The effects of this channel coupling can be quite significant
in the observed cross section, particularly close to 
thresholds \cite{Close:2008hv,Voloshin:2006pz}. 
This was 
studied for $e^+e^-$ annihilation some time ago 
\cite{Eichten:1975ag,Eichten:1978tg,Eichten:1979ms}.  Given the 
observation of new charmonium-like states in channels other than 
$J^{PC}=1^{--}$, it would be useful to revisit these studies. 

\section{Experimental Background}

The observations described below were made
possible by the extraordinary performance of
the PEPII $B$-factory at the Stanford Linear 
Accelerator Center (SLAC) in the U.S., and the KEKB 
$B$-factory at the High Energy Accelerator
Research Organization (KEK) in Japan.  These
$B$-factories, which were constructed to test
the Standard Model  mechanism for matter-antimatter
asymmetries --- so-called $CP$ violation, 
are very high luminosity\footnote{Luminosity
is a measure of the beam-beam interaction intensity.} 
electron ($e^-$) positron ($e^+$) colliders
operating at a center-of-mass (cm) energy near
10,580~MeV.  Electron-positron annihilations at this 
cm energy produce large numbers of $B$-meson anti$B$-meson
($B\bar{B}$) pairs in a coherent quantum 
state.\footnote{A $B$ meson is formed from a $b$ quark and
a $\bar{d}$ or $\bar{u}$ quark in a $1^1S_0$ state.}  
Measurements by
the BaBar experiment~\cite{BaBar_CPV} at PEPII
and the Belle experiment~\cite{Belle_CPV} at KEKB
of the decay patterns of neutral $B^0\bar{B^0}$
pairs have provided sensitive tests of the Standard Model
mechanism for $CP$ violation.

An unexpected bonus from the $B$ factories has been
a number of interesting contributions to the field of
hadron spectroscopy, in particular in the area of
charmonium spectroscopy.  At $B$ factories, charmonium 
mesons are produced in a number of ways.  Here we briefly
describe the charmonium production mechanisms relevant
to the $XYZ$ states.

\paragraph{$B$ decays to final states containing $c\bar{c}$ mesons.} 
$B$ mesons decay radioactively with a lifetime of
approximately 1.5~picoseconds.  At the quark level, the dominant
decay mechanism is the weak-interaction transition of a $b$ quark
to a $c$ quark accompanied by the emission of a 
virtual $W^-$ boson, the mediator of the weak 
interaction.  Approximately half
of the time, the $W^-$ boson materializes as a $s\bar{c}$
pair.  As a result, almost half of all $B$ meson decays
result in a final state that contains a $c$ and $\bar{c}$
quark.  When these final-state $c$ and $\bar{c}$ quarks
are produced close to each other in phase space, they can
coalesce to form a $c\bar{c}$ charmonium meson.

The simplest charmonium-producing $B$-meson decays are those
where the $s$ quark from the $W^-$ combines with the
parent $B$ meson's $\bar{u}$ or $\bar{d}$ quark to
form a $K$ meson.   In such decays, to the extent that the  
$\bar{u}$ or $\bar{d}$ quark act as a passive 
spectator to the decay process, the possible 
$J^{PC}$ quantum numbers of the produced $c\bar{c}$
charmonium system are $0^{-+}$, $1^{--}$ and $1^{++}$.
Experimentally it is observed that decays of
the type $B\rt K(c\bar{c})$, where the $c\bar{c}$ pair
forms a charmonium state with these $J^{PC}$ values,
occur with branching fractions that are all within about
a factor of two of $1\times 10^{-3}$.  Since both of the
$B$ factory experiments detect more than a million
$B$ mesons a day, the number of detected charmonium
states produced via the $B\rt K(c\bar{c})$ process
is substantial.  In 2002, the Belle group discovered
the $\eta^{\prime}_c$, the first radial excitation of
the $\eta_c$ meson, 
via the process 
$B\rt K\eta^{\prime}_c$, where
$\eta^{\prime}_c\rt K_S K^-\pi^+$~\cite{Belle_etac2s}.

\paragraph{Production of $1^{--}$ charmonium states via
initial state radiation.} \footnote{ISR: initial state radiation}
In $e^+ e^-$ collisions at a cm energy of 10,580~MeV,
the initial-state $e^+$ or $e^-$ occasionally radiates
a high-energy $\gamma$-ray, and the  $e^+$ and $e^-$
subsequently annihilate at a correspondingly reduced cm
energy.  When the cm energy of the radiated gamma ray ($\gamma_{ISR}$)
is between 4000 and 5000~MeV, the $e^+e^-$ annihilation
occurs at cm energies that correspond to the range of $mc^2$
values of charmonium mesons.  Thus, the initial state
radiation (ISR) process can directly produce charmonium
states with $J^{PC}=1^{--}$.  Although this is a suppressed 
higher-order QED process, the very high luminosities available
at the $B$ factories have made ISR a valuable 
research tool.   For example, the BaBar group used the ISR
technique to make measurements of $J/\psi$ meson
decay processes~\cite{BaBar_jpsi_2_3pi}.
 
\paragraph{Charmonium associated production with $J/\psi$ mesons in 
$e^+e^-$ annihilation.}
In studies of  $e^+e^-$ annihilations at cm energies near 10,580~MeV, the 
Belle group made the unexpected discovery that when a $J/\psi$ meson
is produced in the inclusive annihilation process 
$e^+e^-\rt J/\psi$+~anything, there is a high probability 
that the accompanying system will
contain another $c\bar{c}$ meson system~\cite{Belle_jpsi_incl}.  
Belle finds, for example, cross sections for the annihilation processes
$e^+e^- \rt J/\psi \eta_c$, $J/\psi \chi_{c0}$ and $J/\psi\eta^{\prime}_c$
that are more than an order-of-magnitude larger than were previously 
expected~\cite{NRQCD}.  Thus, the study of systems recoiling
against the $J/\psi$ in inclusive $e^+e^- \rt J/\psi X$  
annihilations  is another source of $c\bar{c}$ states
at a $B$ factory.   The very low cross sections for these processes
is compensated by the very high luminosities enjoyed by the BaBar
and Belle experiments.
The conservation of charge-conjugation parity in 
electromagnetic  processes guarantees that the $C$ quantum number of the 
accompanying $c\bar{c}$ system will be $C=+$. 
Experimentally, only  the $0^{-+}$
$\eta_c$ and $\eta^{\prime}_c$, and the $0^{++}$ $\chi_{c0}$
are observed to be produced in association with a $J/\psi$. The
$1^{++}$ $\chi_{c1}$ and $2^{++}$ $\chi_{c2}$ are not 
seen.
This indicates that this process favors
the production of $J=0$ states over those with $J=1$ and 
higher.

\paragraph{Two-photon collisions}
In high energy $e^+e^-$ machines,  photon-photon 
collisions are produced when both an incoming
$e^+$ and $e^-$  radiate  photons that subsequently
interact with each other.  Two-photon interactions
can directly produce particles with 
$J^{PC}=0^{-+}$, $0^{++}$, $2^{-+}$ and $2^{++}$.  
An example is the CLEO group's confirmation of
the existence of the $\eta^{\prime}_c$ charmonium state
from studies of two-photon production of $K_SK^{\pm}\pi^{\mp}$
final states~\cite{CLEO_etac2s}.

\section{Experimental evidence and theoretical interpretations for the 
$XYZ$ mesons}

In this section we describe experimental
characteristics of the $XYZ$ mesons, discuss their various theoretical 
interpretations,  and present measurements that can distinguish 
between possibilities and verify their nature.

\subsection{$X(3872)$}

The $X(3872)$ was first seen in 2003 by Belle 
as a  narrow peak near $mc^2= 3872$~MeV in the
$\pi^+\pi^- J/\psi$ invariant mass distribution
in $B^-\rt K^-\pi^+\pi^- J/\psi$ decays~\cite{Belle_x3872}.
Shortly after the Belle announcement, it was observed by
the CDF~\cite{CDF_x3872} and D0~\cite{D0_x3872} groups to be produced in 
high energy
proton-antiproton ($p\bar{p}$) collisions at the Fermilab 
Tevatron, and its production in $B$ meson decays was 
subsequently confirmed by BaBar~\cite{BaBar_x3872}.  The current 
world average 
mass is $(3871.4 \pm 0.6)$~MeV/$c^2$, and its total width is
less than 2.3~MeV/$c^2$~\cite{PDG}.   

%
Both BaBar \cite{Aubert:2006aj} and Belle  \cite{Abe:2005ix} have reported
evidence for the 
decay $X(3872)\to \gamma J/\psi$, which indicates that the $X(3872)$
has $C=+$.  This implies that the dipion in the $\pi^+\pi^-J/\psi$ 
mode has $C=-$, suggesting it originates from a $\rho$.
In fact, analyses of 
$X(3872)\rt\pi^+\pi^- J/\psi$ decays by the CDF group
have demonstrated that the dipion  invariant mass distribution 
is most simply understood by the hypothesis that
it originates from the 
decay $\rho\rt\pi^+\pi^-$~\cite{CDF_rho}.
The decay of a charmonium state to $\rho J/\psi$
would violate isospin, and the evidence for this process provides a strong
argument against a charmonium explanation for this state.
Evidence for the  decay $X(3872) \to \pi^+ \pi^- \pi^0 J/\psi$ 
at a rate comparable to that of $\pi^+ \pi^- J/\psi$ was reported by
Belle~\cite{Abe:2005ix} leading to speculation that the decay proceeds 
through a virtual $\omega$, as had been predicted by Swanson~\cite{Swanson:2003tb}.
If confirmed, the co-existence of both the $X(3872) \to \pi^+
\pi^- J/\psi$ and $X(3872) \to \pi^+ \pi^- \pi^0 J/\psi$ transitions implies
that the $X(3872)$ is a mixture of both $I=0$ and $I=1$, 
as suggested by Close and Page~\cite{Close:2003sg}.
Angular correlations among the final state
particles from $X(3872)\rt\pi^+\pi^- J/\psi$ decays
rule out all $J^{PC}$ assignments for
the $X(3872)$ other than $J^{PC}=1^{++}$ 
and $2^{-+}$~\cite{CDF-Jpc}.   Neither of the available
charmonium assignments with these $J^{PC}$ values:
the $1^{++}$ $\chi^{\prime}_{c1}$  (the $2^3P_1$ $c\bar{c}$ state) 
or the $2^{-+}$ $\eta_{c2}$  (the $1^1D_1$ $c\bar{c}$ state),
are expected to have large branching fractions for the 
isospin-violating $\rho J/\psi$ decay channel.
Furthermore, a $2^{-+}$ assignment would require that
the decay $X(3872)\to \gamma J/\psi$ be a highly 
suppressed higher multipole, and therefore unlikely. 
Finally, in the discussion below we identify the $Z(3930)$ as the $2^3P_2$ state,
thereby setting the $2^3P_2$ mass at $\sim 3930$~MeV/$c^2$; this is inconsistent with the 
$2^3P_1$ interpretation of the $X(3872)$, since the $2^3P_2 - 2^3P_1$ mass splittings
are expected to be lower than $\sim$50~MeV/$c^2$~\cite{Eichten:2005ga,Barnes:2005pb}. 
 
An intriguing feature
of the $X(3872)$ is that its measured mass value is very
nearly equal to the sum of the masses of the $D^0$ and
$D^{*0}$ mesons,
which has recently been precisely measured by the CLEO 
experiment to be
$m_{D^0} + m_{D^{*0}} = (3871.81\pm 0.36$)~MeV/$c^2$~\cite{CLEO_mD0}.
This close correspondence has led to considerable speculation
that the $X(3872)$ is a molecule-like bound state of
a $D^0$ and  a $\bar{D^{*0}}$ meson
\cite{Voloshin:1976ap,DeRujula:1976qd,tornqvist_x3872,Swanson:2003tb,
Close:2003sg,Voloshin:2003nt}.\footnote{In this review the inclusion of charge conjugate states
is always implied. {\it E.g.,} ``a $D^0$ and  a $\bar{D^{*0}}$ meson''
could also mean a $D^{*0}$ and  a $\bar{D^{0}}$ meson.}
The recent measurement of the $D^0$ mass implies a
$D^{*0}\bar{D}^0$ binding energy of $(0.6\pm 0.6)$~MeV/$c^2$. 

A $1^{++}$ quantum number assignment for the $X(3872)$ implies that  $S$-wave couplings of the
$X$ to $D^{*0}\bar{D}^0$ is permitted and these 
result in a strong coupling between the $X$ 
and the two mesons.  This strong coupling can produce a bound state 
with a molecular structure just below
the two-particle threshold.  In this molecular scenario, the decays of the $X(3872)$ into
$D^{*0}\bar{D}^0\pi^0$ and $D^{*}\bar{D}^0\gamma$ proceed through the decays of its constituent
$D^{*0}$ with branching ratios similar to those of the 
$D^{*0}$~\cite{Voloshin:2003nt,Close:2003sg}. 
The $D^0\bar{D}^{0*}$ molecule wavefunction is expected to contain some admixture of
$\rho J/\psi$ and $\omega J/\psi$.  This explains the isospin-violating $\rho J/\psi$ decay 
mode and also successfully predicted the $\pi^+\pi^- \pi^- J/\psi$ decay width 
\cite{Swanson:2003tb}.
A further prediction of the molecule explanation is 
the existence of a molecular $D^*\bar{D}^*$ state with mass 4019~MeV/$c^2$ and 
$J^{PC}=0^{++}$
decaying to $\omega J/\psi$, $\eta\eta_c$ and $\eta'\eta_c$ \cite{Swanson:2006st}. 
A related possibility is that the $X(3872)$ is dynamically generated~\cite{Gamermann:2007fi}.

Maiani {\it et al} advocate a tetraquark explanation for the $X(3872)$
\cite{Maiani:2004vq,Ebert:2005nc}.  A prediction of the tetraquark interpretation
is the existence of a second neutral $X$ state
that is strongly produced in neutral $B^0$ mesons decays 
to $K^0 \pi^+\pi^- J/\psi$
with a mass  that differs by $8\pm3$~MeV/$c^2$
from the mass of the $X(3872)$ produced in $B^+$
decays.  They also predicted
the existence of charged isospin partner states.

\begin{figure}[htb!]
\centerline{\epsfxsize 4.0 truein
\epsfbox{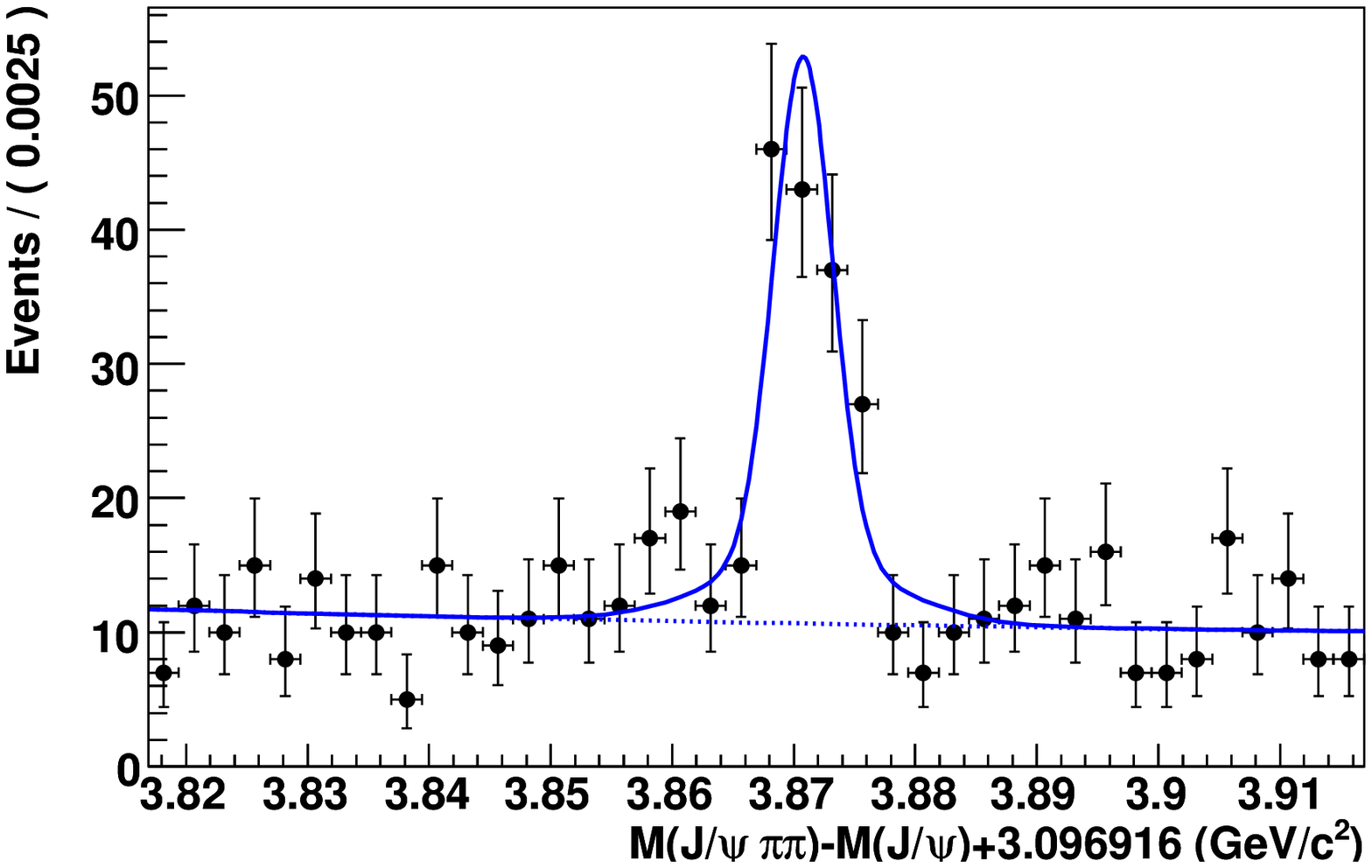}}
   \vspace{1mm}
\centerline{\epsfxsize 4.0 truein
\epsfbox{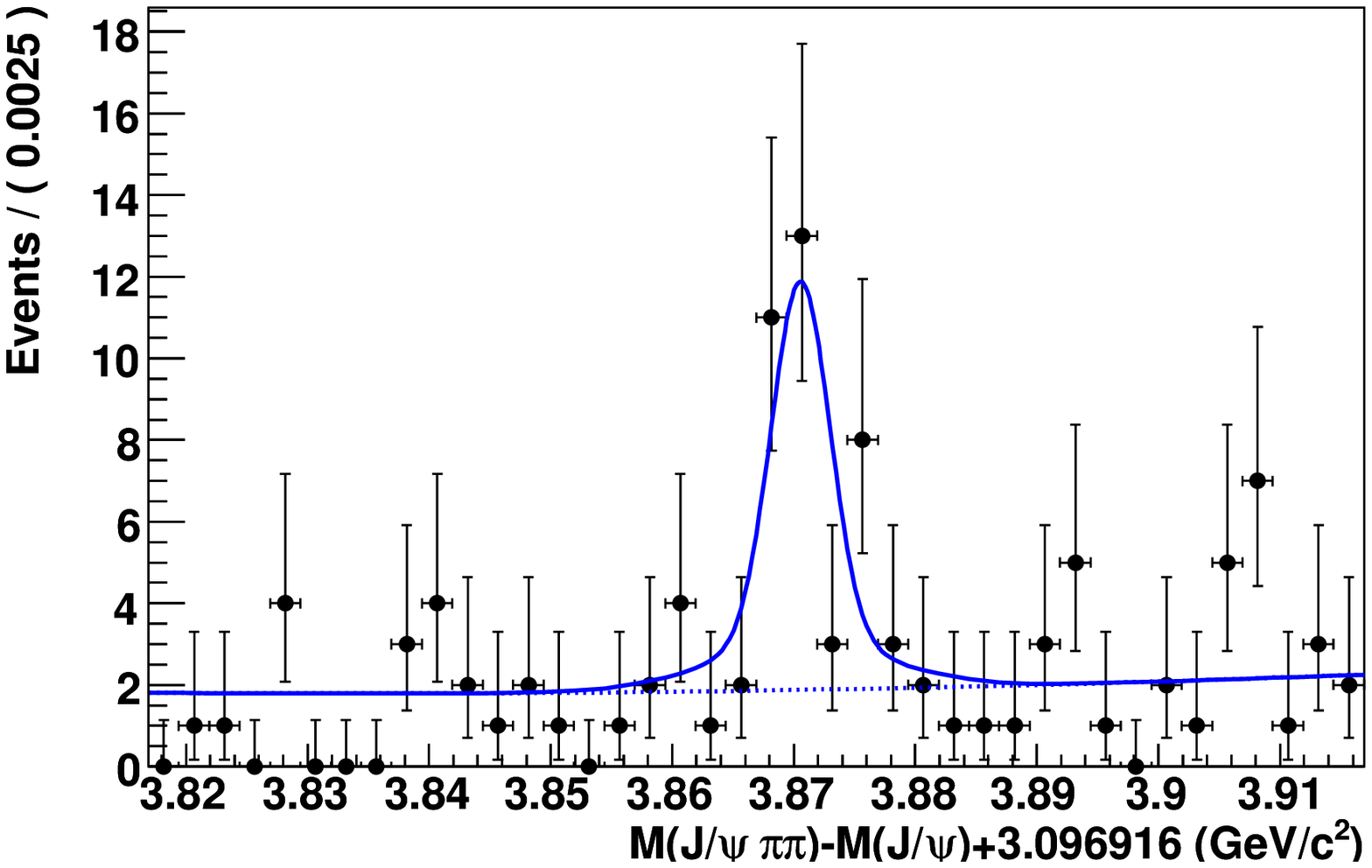 }}
\caption{\label{fig:Belle_x3872}
The $M(\pi^+ \pi^- J/\psi)$ mass distributions produced in charge $B^-$
(top) and neutral $B^0$ (bottom) decays to $K \pi^+\pi^- J/\psi$ final 
states~\cite{Belle_x38720}.
}
\end{figure}

Belle recently reported a $\sim 7\sigma$ signal
for $X(3872)$ production in neutral $B$ meson
decays ({\it i.e.}, $B^0\rt K_S^0 X(3872)$),
which is shown, together with the signal from
charged $B$ meson decays, in Fig.~\ref{fig:Belle_x3872}.
They determine a mass difference between the $X(3872)$ produced 
in charged  versus neutral B decays of 
$0.9\pm 0.9$~MeV/$c^2$~\cite{Belle_x38720},
which is consistent with zero, and inconsistent with
the tetraquark-model prediction.  This confirms an
early lower-precision result from BaBar~\cite{BaBar_x38720}.

The BaBar group also
searched for a charged partner of the $X(3872)$ in the $\rho^{-} J/\psi$
decay channel.  They found no signal and excluded
an isovector hypothesis for the $X(3872)$ with high 
confidence~\cite{BaBar_xminus3872}.

In studies of the decay process $B\rt K D^0\bar{D^0}\pi^0$,
both Belle~\cite{Belle_DDpi} and BaBar~\cite{BaBar_DDpi}
report narrow enhancements just above the $m_{\pi^0}+2m_{D^0}$
mass threshold.  The peak mass values for the two observations
($3875.2\pm 1.9)$~MeV/$c^2$ for Belle and $(3875.1\pm 1.2)$~MeV/$c^2$ 
for BaBar) are in good agreement with each other and about 
four standard deviations higher 
than the $X(3872)$ mass measured with $\pi^+\pi^- J/\psi$ decays,
which may be evidence for the second $X$ state predicted in Ref.~\cite{Maiani:2004vq}.
BaBar observes this enhancement in both the $D^0\bar{D}^0\pi^0$ and $D^0\bar{D}^0\gamma$
modes, which strongly supports the presence of the $D^0\bar{D}^{*0}$ intermediate state in
the decay of this (possibly) new $X$.

Dunwoodie and Ziegler argue that the mass shift between the two sets of observations
is a result of the sensitivity of the peak position of the $D^*\bar{D}^0$ invariant mass 
distribution to the final state orbital momentum because of the proximity of the $X(3872)$ 
to the $D^*\bar{D}^0$ threshold \cite{Dunwoodie:2007be}.  Another explanation 
put forward in Refs.~\cite{Hanhart:2007},~\cite{Voloshin:2007hh} and~\cite{Braaten:2007ft}  is 
that the line shapes of the $X(3872)$
depend on its decay channel and are different for the $J/\psi\pi^+\pi^-$,
$J/\psi\pi^+\pi^-\pi^0$, and $D^0\bar{D}^0\pi^0$ channels. In both explanations the more
massive enhancement is not regarded as a separate state, but a manifestation of
the $X(3872)$.

To summarize, there is an emerging consensus that the $X(3872)$ is a multiquark state with 
the molecular interpretation being favoured due to its proximity to the $DD^*$ threshold.

\subsection{$XYZ$ particles with $mc^2$ near 3940~MeV}

Three, apparently distinct, $XYZ$ candidate states have 
been observed with 
masses near 
3940~MeV/$c^2$.  These include the $X(3940)$, seen as a peak
in the $D\bar{D^*}$ invariant mass in the process 
$e^+e^-\rt J/\psi D\bar{D^*}$~\cite{Belle_x3940};
the $Y(3940)$, a peak in the $\omega J/\psi$ mass spectrum seen
in $B\rt K \omega J/\psi$ decays~\cite{Belle_y3940}; and the
$Z(3930)$, a peak in the invariant mass distribution of $D\bar{D}$ 
meson  pairs produced in two-photon collisions~\cite{Belle_z3930}.

\subsubsection{The $Z(3930)$}

\begin{figure}[htb!]
\begin{center}
\centerline{\epsfig{file=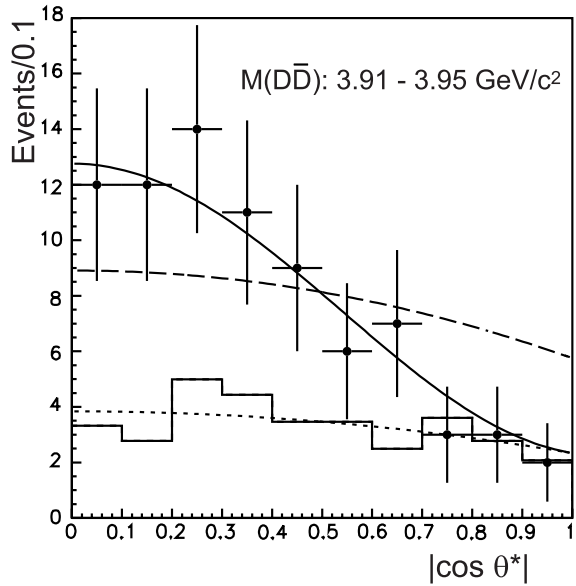,width=3.0in,clip=}}
\caption{Belle's $\chi_{c2}(2P)$ candidate~\cite{Belle_z3930}:
$\cos \theta^*$, the angle of the $D$~meson relative to the 
beam axis in the $\gamma\gamma$ center-of-mass frame for
events with $3.91 < m(D \bar D) < 3.95\,\mbox{GeV}$; the data (circles) are
compared with predictions for $J=2$ (solid) and $J=0$ (dashed). 
The background level can be judged from the solid histogram
or the interpolated smooth dotted curve.
\label{fig:belle_z3930}}
\end{center}
\end{figure}

The $Z(3930)$ seems easiest to understand.   
It is a peak reported by Belle in the spectrum of $D\bar{D}$ mesons produced 
in $\gamma\gamma$ collisions, with  mass and width 
$M=3929\pm 6$~MeV/$c^2$ and 
$\Gamma=29\pm 10 $ MeV/$c^2$~\cite{Belle_z3930}.
The $D \bar D$ decay mode makes it impossible
for the $Z(3930)$ to be the $\eta_c(3S)$ state. 
The two-photon production process can only produce $D\bar{D}$ 
in a $0^{++}$ or $2^{++}$ state and for these, the $dN/d\cos\theta^*$
distribution, where $\theta^*$ is the angle between the $D$ meson 
and the incoming photon in the $\gamma\gamma$ cm, are quite distinct:  
flat for $0^{++}$ and $\propto \sin^4\theta^*$ for $2^{++}$.  The
Belle measurement strongly favors the $2^{++}$ hypothesis
(see Fig.~\ref{fig:belle_z3930}), making 
the $Z(3930)$
a prime candidate for the $\chi^{\prime}_{c2}$, the $2^3P_2$ charmonium state.
The predicted mass of the $\chi_{c2}(2P)$ is 3972~MeV/$c^2$ and the
predicted total width assuming the observed mass value is
$\Gamma_{\rm 
total}(\chi_{c2}(2P))=28.6$~MeV/$c^2$~\cite{swanson,Barnes:2005pb,Eichten:2005ga},
in good agreement with the experimental measurement.  Furthermore, 
the two-photon production rate for
the $Z(3930)$ is also consistent with expectations for 
the $\chi^{\prime}_{c2}$~\cite{Barnes:1992sg}.
The $\chi_{c2}(2P)$ interpretation could be confirmed by observation of the
$D\bar{D}^*$ final state, which is expected to have a 
${\cal B} \sim 25\%$~\cite{Eichten:2005ga,Barnes:2005pb}, and  the radiative transition 
$\chi_{c2}(2P)\to \gamma \psi(2S)$ which is predicted to have a 
partial width of ${\cal O}(100\,\mathrm{keV})$~\cite{Eichten:2005ga,Barnes:2005pb}.

\subsubsection{The $X(3940)$ (and $X(4160)$)}

Belle observed the $X(3940)$ in double-charmonium production in the reaction 
$e^+e^-\to J/\psi +X$ with mass $M=3943 \pm 8$~MeV/c$^2$ and intrinsic 
width $\Gamma< 52$~MeV/c$^2$ at the 90\% C.L.~\cite{Belle_x3940}.  
In addition to the $X(3940)$, Belle observed the well known
charmonium states $\eta_c$, $\chi_{c0}$, and $\eta_c(2S)$ with properties consistent
with those from other determinations. 
While a distinct signal for $X(3940)\rt D\bar{D^*}$ is seen, 
there is no evidence for the $X(3940)$ in either the $D\bar{D}$ or $\omega J/\psi$
decay channels.  If the $X(3940)$ has $J=0$, as seems to be the case
for mesons produced via this production mechanism, 
the absence of a substantial
$D\bar{D}$ decay mode strongly favors $J^{P}=0^{-+}$, 
for which the most likely charmonium assignment 
is the $\eta^{\prime\prime}_c$, the $3^1S_1$ charmonium
state.  
The fact that the lower mass $\eta_c(1S)$ and $\eta_c(2S)$ are also produced in
double charm production seems to support this assignment.
The predicted width for a $3^1S_0$ state with a mass of 3943~MeV/$c^2$ is 
$\sim 50$~MeV/$c^2$~\cite{Eichten:2005ga}, which is in acceptable
agreement with the measured $X(3940)$ width.  

However, there are problems with this assignment, the first being 
that the measured mass
of the $X(3940)$, recently updated by Belle to be
$(3942\pm 8)$~MeV/$c^2$~\cite{Belle_x4160}, is  below potential model
estimates of $\sim$4050~MeV/$c^2$ or higher~\cite{barnes_0505002}.  
A further complication is the recent observation by
Belle of a mass peak in the $D^*\bar{D^*}$ system
recoiling from a $J/\psi$ in the process $e^+e^-\rt J/\psi 
D^*\bar{D^*}$~\cite{Belle_x4160}.  This state, designated as
$X(4160)$, has a mass of $(4156\pm 29)$~MeV/$c^2$ and a
total width of $\Gamma = (139^{+113}_{-65})$~MeV/$c^2$. 
Using similar arguments, this latter state could also
be attributed to  the $3^1S_0$ state.
But the $X(4160)$ mass is well above expectations for the $3^1S_0$
and well below those for the $4^1S_0$, which is predicted to be
near 4400~MeV/$c^2$~\cite{barnes_0505002}.  
Although {\it either} the $X(3940)$ {\it or} the $X(4160)$ might conceivably fit a 
charmonium assignment, it seems very unlikely that both of them could be 
accommodated as $c\bar{c}$ states.
The $\eta_c^{\prime\prime}$ assignment can be tested by studying
the angular distribution of the $D\bar{D}^*$ final state and to 
observe it in $\gamma\gamma \to D\bar{D}^*$.

\subsubsection{The $Y(3940)$}

Belle's observation of the $Y(3940)\rt \omega J/\psi$ 
in $B \rt K \omega J/\psi$ decays~\cite{Belle_y3940}  has recently
been confirmed by BaBar~\cite{BaBar_y3940}.
Belle reports a mass and width of $M=(3943\pm 17)$~MeV/c$^2$ and
$\Gamma=(87\pm 34)$~MeV/$c^2$ while BaBar reports the preliminary values 
of
$M=(3914.3^{+4.1}_{-3.8})$~MeV/$c^2$ and 
$\Gamma=(33^{+12}_{-8})$~MeV/$c^2$, which are both somewhat smaller than 
Belle's values.
The measured product branching fractions agree:
${\cal B}(B\rt KY(3940)){\cal B}(X(3940)\rt\omega J/\psi) 
= (7.1 \pm 3.4)\times 10^{-5}$ (Belle), and 
${\cal B}(B\rt KY(3940)){\cal B}(X(3940)\rt\omega J/\psi) 
= (4.9 \pm 1.1)\times 10^{-5}$ (BaBar).
These values, together with an assumption that the
branching fraction ${\cal B}(B\rt K Y(3940)$ is less than
or equal to $1\times 10^{-3}$, a value that is typical
for allowed $B\rt K$+charmonium decays,  imply 
a partial width $\Gamma(Y(3940)\rt \omega J/\psi) > 1$~MeV/$c^2$, 
which is at least an order-of-magnitude higher
than those for hadronic transitions between any of the
established charmonium states.  The Belle group's 90\% 
confidence level limit on 
${\cal B}(X(3940)\rt\omega J/\psi) < 26\%$~\cite{Belle_x3940}
is not stringent enough to rule out the possibility that
the $X(3940)$ and the $Y(3940)$ are the same state.

The mass and width of $Y(3940)$ suggest a radially excited $P$-wave charmonium state.  
We expect that $\chi_{c1}(2P) \rt D\bar{D}^*$ 
would be the dominant decay mode, with a predicted partial
width of 140~MeV/$c^2$ \cite{Barnes:2006xq}, which is consistent with the width of the
$Y(3940)$ within the theoretical and experimental uncertainties.  Furthermore,
the $\chi_{c1}$ is also seen in $B$-decays.  However, 
the large branching fraction for  $Y\to \omega J/\psi$
is unusual for a $c\bar{c}$ state above open charm
threshold.   A possible
explanation for this unusual decay mode is that rescattering through
$D\bar{D}^*$ is responsible: $1^{++} \to D\bar{D}^*\to \omega J/\psi$.  
Another contributing factor might be mixing with the  
molecular state that is tentatively identified with the $X(3872)$.  
The $\chi_{c1}(2P)$ assignment can be tested by searching for the $D\bar{D}$
and $D\bar{D}^*$ final states and by studying their angular distributions. With
the present experimental data, a $\chi_{c0}(2P)$ assignment
cannot be ruled out.

\subsection{The $J^{PC}=1^{--}$ states produced via ISR}

Using the ISR process, BaBar discovered unexpected
peaks near 4300~MeV/$c^2$ in the $\pi^+\pi^- J/\psi$ and 
$\pi^+\pi^-\psi^{\prime}$ channels.  The partial
widths for these decay channels are much 
larger than
usual for charmonium states.  Since these states
are produced via the ISR process, they have
$J^{PC}=1^{--}$.  All of the $1^{--}$ charmonium levels 
in the $4000$-$4500$~MeV/$c^2$ mass range have already been assigned
to the well established $\psi(4040)$, $\psi(4160)$ 
and $\psi(4415)$ mesons, which match well their assignments
to the $3^3S_1$,  $2^3D_1$ and $4^3S_1$ $c\bar{c}$ states, respectively. 
The quark model predicts additional charmonium $1^{--}$ states, 
but at higher masses;  $3^3D_1(4520)$, $5^3S_1(4760)$,  $4^3D_1(4810)$,
etc. \cite{Godfrey:1985xj}.
The experimental situation is likely complicated and obscured by interference between
these resonances which can affect the parameters extracted from the
measurements.

\subsubsection{The $Y(4260)$ $\pi^+\pi^- J/\psi$ resonance}
The BaBar group measured the energy dependence of the
cross section for $e^+e^- \rt \pi^+\pi^- /\psi$ using
ISR radiation events at a primary cm energy of 10,580~MeV.
They found a broad enhancement around 4260~MeV~\cite{BaBar_y4260}
that they dubbed the $Y(4260)$.
A fit to the peak with a single Breit Wigner resonance shape
yields a mass $M=(4259\pm 10)$~MeV/$c^2$ and full width 
$\Gamma=(88 \pm 24)$~MeV/$c^2$, values that
are quite distinct from those
of other established charmonium states.  Although it is well
above  the threshold for decaying into $D\bar{D}$, $D\bar{D^{*}}$
or $D^{*}\bar{D^{*}}$ meson pairs, there is no evidence for the
$Y(4260)$ in any of these channels~\cite{Belle_galina}.  In fact,
there appears to be a dip in the measured
 $e^+e^-$ total annihilation cross 
section at this energy~\cite{BES_R}.  An
analysis using total cross section data for $e^+e^-$ annihilation 
into hadrons at cm energies around 4260~MeV results in a 
90\% CL lower limit on the partial decay width 
$\Gamma(Y(4260)\rt\pi^+\pi^- J/\psi)>1.6$~MeV/$c^2$~\cite{Mo_y4260}, 
which is much 
larger than the partial widths for equivalent transitions among established
$1^{--}$ charmonium states.  The $Y(4260)$ peak was confirmed by both
CLEO~\cite{CLEO_y4260} and Belle~\cite{Belle_y4260}. 
Belle reported a second, broader $\pi^+\pi^- J/\psi$ peak 
near 4008~MeV/$c^2$.   It is currently not known whether this
latter enhancement is due to the $\psi(4040)$, 
a dynamical threshold enhancement, or another meson state.

\subsubsection{The $\pi^+\pi^- \psi'$ resonances at 4370~MeV/$c^2$ and 
4660~MeV/$c^2$}

\begin{figure}[htb!]
\centerline{\epsfysize 3.0 truein
\epsfbox{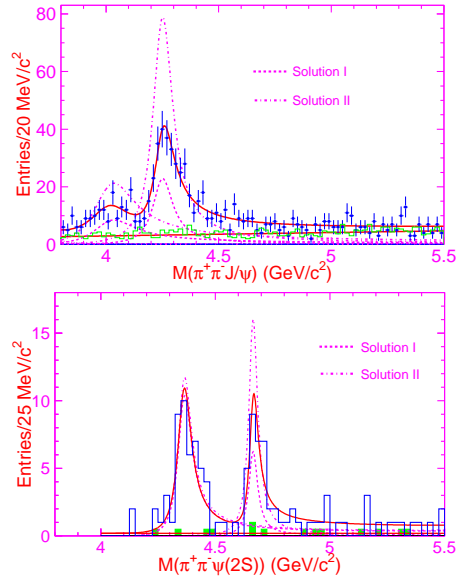 }}
\caption{The $\pi^+ \pi^- J/\psi$ (Top) and
$\pi^+\pi^- \psi^{\prime}$ (Bottom) invariant mass
distributions for the ISR processes 
$e^+ e^- \rt  \gamma \pi^+ \pi^- J/\psi (\psi^{\prime})$,
from Refs.~\cite{Belle_y4260,Belle_y4660}.
The curves indicate the results of fits of interfering
Breit Wigner resonances to the data.
}
\label{fig:y4260_etc}
\end{figure}

BaBar also found a broad peak in the cross section for $e^+e^-\rt 
\pi^+\pi^- \psi'$ that is distinct from the $Y(4260)$; its peak 
position and width are not consistent with those of the 
$Y(4260)$~\cite{BaBar_y4325}.  The BaBar observation was 
subsequently confirmed by a Belle study that 
used a larger data sample~\cite{Belle_y4660}.   
Belle found
that the $\pi^+\pi^-\psi'$ mass enhancement is, in fact, produced by
two distinct peaks, one, the $Y(4360)$ with $M= (4361 \pm 13)$~MeV/$c^2$ 
and $\Gamma = (74 \pm 18)$~MeV/$c^2$ and a second, the $Y(4660)$ with 
$M=(4664 \pm 12)$~MeV/$c^2$
and $\Gamma = (48 \pm 15)$~MeV/$c^2$~\cite{Belle_y4660}.  These
masses and widths are not consistent with any of the 
established $1^{--}$ charmonium states, and no sign
of a peak at either of these masses is evident either in the
$e^+e^-$ total annihilation cross section~\cite{BES_R} 
or in the exclusive cross sections  $e^+e^- \rt D\bar{D}$,
$D\bar{D^*}$, or $D^*\bar{D^*}$~\cite{Belle_galina}, which
indicates that the $\pi^+\pi^- \psi'$ partial width for these
states is unusually large (at least by charmonium standards). 
Moreover, as is evident in Fig.~\ref{fig:y4260_etc}, which
shows the recent Belle results for $\pi^+\pi^- J/\psi$ (top)
and $\pi^+\pi^- \psi^{\prime}$ (bottom) with the same 
horizontal mass scales,
there is no sign of either the $Y(4360$ or $Y(4660)$ in the 
$\pi^+\pi^- J/\psi$ channel; nor is there any sign of
the $Y(4260)$ peak in the $\pi^+\pi^- \psi^{\prime}$ mass spectrum.

\subsubsection{Discussion}

The discovery of the $Y(4260)$,  $Y(4360)$, and  $Y(4660)$
appears to represent an overpopulation 
of the expected charmonium $1^{--}$ states. 
The absence of open charm production is also inconsistent with
a conventional $c\bar{c}$ explanation. 
While Ding {\it et al} argue that the $Y(4360)$ and $Y(4660)$ are conventional 
$c\bar{c}$ states, in particular the $3^3D_1$ and $5^3S_1$, 
respectively~\cite{Ding:2007rg}, their masses are inconsistent with other 
quark model calculations~\cite{Godfrey:1985xj}. 
Other possible explanations of these state include charmonium hybrids, 
$S$-wave charm meson thresholds, or multiquark states; either 
$cq\bar{c}\bar{q}$ tetraquarks or $DD_1$ and $D^*D_0$ molecules. 
Liu suggests that the peak at 4008~MeV/c$^2$ is related to the $D^*\bar{D}^*$ threshold
and could be 
a $D^*\bar{D}^*$ molecule where the $D^*$ and $\bar{D}^*$ are in a 
$P$-wave~\cite{Liu:2007ez}.

The $Y(4260)$ has been around the longest and has received the most scrutiny.  
The first unaccounted-for $c\bar{c}$  
state is the $\psi(3^3D_1)$ which is predicted to have a mass of
$M(3^3D_1) \simeq 4500$~MeV/c$^2$, much too heavy to be the $Y(4260)$.  
Numerous explanations have been proposed: 
it is a $D^*(2010)\bar{D}_1(2420)$ threshold enhancement \cite{Rosner:2006vc}, 
a $D \bar D_1$ or $D^*\bar{D}_0^0$ bound state~\cite{y4260_as_DD1,Rosner:2006sv,y4260_as_dstr-d2str},
or a $c s \bar c \bar s$ tetraquark~\cite{Maiani:2005pe},
In the latter
cases, the $Y$ would decay to $D \pi \bar D^*$, where the $D$ and $\pi$ are not from a 
$D^*$.  
One would expect this mode to have a large width so its non-observation 
disfavours the tetraquark/molecule explanations.   

An attractive 
interpretation is that the $Y(4260)$ is a 
charmonium hybrid \cite{Zhu:2005hp,Close:2005iz,Kou:2005gt}.  
The flux tube model predicts the lowest $c\bar{c}$ hybrid to have a 
mass that is $\sim 4200$~MeV/$c^2$~\cite{Barnes:1995hc}, and this is
consistent with lattice gauge theory predictions~\cite{Lacock:1996ny}.  
Lattice gauge theory found that the $b\bar{b}$ hybrids have
large couplings to closed flavor
channels \cite{McNeile:2002az} which is similar
to the BaBar observation of $Y\to J/\psi \pi^+\pi^-$ and is much larger
than is typical for transitions involving conventional charmonium states. 
A prediction of the hybrid hypothesis is that the dominant 
hybrid-charmonium open-charm decay modes are expected to be a meson pair with an
$S$-wave ($D$, $D^*$, $D_s$, $D_s^*$) and a $P$-wave ($D_J$, $D_{sJ}$) in the
final state~\cite{Close:2005iz}.  The dominant decay mode is expected to be
$D \bar D_1$.  
Evidence for a large $D \bar D_1$ signal would be strong evidence for the
hybrid interpretation.  A complication is that the $D \bar D_1$ threshold is 
4287~MeV/$c^2$ if we consider the lightest $D_1$ to be the narrow state 
at 2422 MeV/$c^2$~\cite{PDG}.  
Note that both the $Y(4370)$ and the $Y(4660)$ are well above
the $D\bar{D_1}$ mass threshold.  If the same hybrid interpretation
is applied to them, decays to $D\bar{D_1}$ should be very strong
and one would expect peaks in the exclusive cross sections for
$e^+e^-\rt D\bar{D_1}$ at these masses.

Lattice gauge theory also suggests that we search for other
closed charm modes with
$J^{PC}=1^{--}$: $J/\psi \eta$, $J/\psi \eta'$, $\chi_{cJ} \omega$ and more.
If the $Y(4260)$ is a hybrid it is expected to be a member of a multiplet
consisting of eight states with masses in the 4000 to 4500~MeV/$c^2$ mass range.
It would be most convincing if some of these partners were found, especially
those with exotic $J^{PC}$ quantum numbers.  In the flux-tube model the exotic states have
$J^{PC}=0^{+-}$, $1^{-+}$, and $2^{+-}$ while the non-exotic low-lying hybrids
have $0^{-+}$, $1^{+-}$, $2^{-+}$, $1^{++}$, and $1^{--}$.

The current situation regarding the $1^{--}$ states produced via ISR 
is clearly unsettled.  
Coupled-channel effects and rescattering of pairs of charmed mesons could play an important
role \cite{Voloshin:2006pz} in understanding these peaks.
Further complications that need to be understood are  couplings to channels 
near thresholds that can interfere with conventional $c\bar{c}$ states.  The challenge
in extracting resonance parameters in this environment is highlighted by the recent
BES analysis which attempted to take into account the interference between these 
broad resonances and found substantial variations in the resonance parameters compared
to a fit that didn't take into account interference~\cite{Ablikim:2007gd}.
Clearly,  more experimental information  on the decay properties of these
states and more theoretical work on coupled channel and interference effects 
are needed if a better understanding of these states is to be achieved.

\subsection{The $Z^+(4430)\rt\pi^+ \psi^{\prime}$}

\begin{figure}[htb!]
\centerline{\epsfysize 3.0 truein
\epsfbox{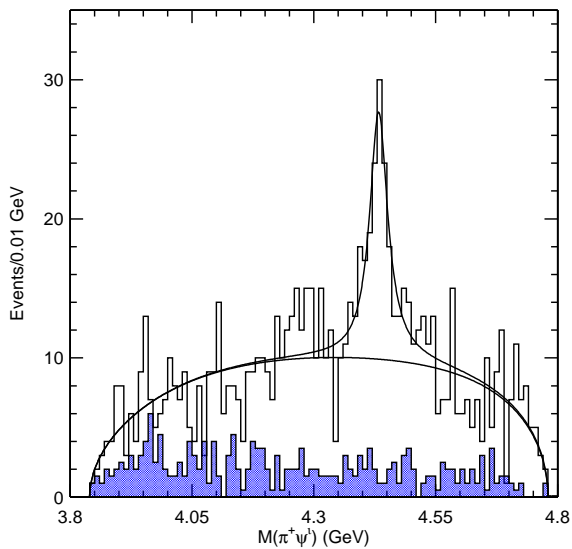 }}
\caption{The $\pi^+ \psi^{\prime}$ invariant mass
distribution for $B\rt  K\pi^+ \psi^{\prime}$ decays~\cite{Belle_z4430}.
The shaded histogram is the estimated background. The
curve is the result of a fit described in the text.
}
\label{fig:z4430}
\end{figure}

In Summer 2007, the Belle group reported on a study of the
$B\rt K\pi^+\psi'$ decay process where they 
observed a relatively narrow enhancement in the $\pi^+ \psi^{\prime}$ 
invariant mass 
distribution at $M=(4433\pm 5)$~MeV/$c^2$~\cite{Belle_z4430}
(see Fig.~\ref{fig:z4430}).
A fit with a single relativistic Breit Wigner, indicated by the
smooth curve in Fig.~\ref{fig:z4430}, yields 
a total width of $\Gamma=(45^{+35}_{-18})$~MeV/$c^2$, 
which is too narrow to 
be caused by interference effects in the $K\pi$ channel.  The
$B$ meson decay rate to this state, which is called $Z^+(4430)$,
is similar to that for decays to the $X(3872)$ and $Y(3940)$, which
implies that the $Z^+(4430)$ has a substantial branching fraction 
({\it i.e.} greater than a few percent) to $\pi^+\psi^{\prime}$ and,
thus, a partial decay width for this mode that is on the MeV scale.  
There are no reports of a $Z^+(4430)$ signal in the $\pi^+J/\psi$
decay channel.

Among the $XYZ$ exotic meson candidates, the $Z^+(4430)$ is unique
in that it has a non-zero electric charge, a feature that cannot
occur for $c\bar{c}$ charmonium states or $c\bar{c}$-gluon hybrid
mesons.  It is, therefore, a prime candidate for a multiquark meson.

There have been a number of theoretical explanations.  Because it is close
to the $D^*\bar{D}_1(2420)$ threshold, Rosner suggested it is an $S$-wave 
threshold effect~\cite{Rosner:2007mu}, while others consider it
to be a strong candidate 
for a $D^*\bar{D}_1(2420)$ molecule~\cite{Meng:2007fu,Lee:2007gs,Liu:2007bf}. 
Maiani {\it et. al.} suggested that the $Z(4430)$ is a diquark-diantiquark state 
with flavour $[cu][\bar{c}\bar{d}]$ and is the radial excitation of a 
$X_{u\bar{d}}^+(1^{+-}; 1S)$ state with mass 3880~MeV/c$^2$~\cite{Maiani:2007wz}.  
The tetraquark hypothesis implies that the $Z(4433)^+$ will have neutral partners
decaying to $\psi(2S) +\pi^0/\eta$ or $\eta_c(2S)+\rho^0/\omega$.
If the $X^+(4430)$ is a molecule, assuming that the $D^*\bar{D_1}$ is in a relative $S$-wave,
it will have $J^P=0^-$, $1^-$, or $2^-$, with the lightest state expected to be
the $0^-$ \cite{Liu:2007bf}.  In contrast, a 
tetraquark would have $J^P=1^+$. The molecule will decay via the decay of
its constituent mesons into $D^*D^*\pi$~\cite{Meng:2007fu}
while the tetraquark will fall apart into 
$D\bar{D^*}$, $D^*\bar{D^*}$, $J/\psi \pi$, $J/\psi \rho$, $\eta_c\rho$ and $\psi(2S)\pi$,
but not into $D\bar{D}$ due to its unnatural spin-parity~\cite{Ding:2007ar}. 
The tetraquark model also predicts a 
second nearby state with mass $\sim$4340~MeV/c$^2$
also decaying into the $\psi' \pi^+$ final state~\cite{Maiani:2007wz}.

\subsection{Are there corresponding states in the $s$ and
$b$ quark sectors?}

Many of the models proposed to explain the $XYZ$ states 
predict analogous states in the $b\bar{b}$
and $s\bar{s}$ sectors.
In the $s\bar{s}$ sector the $f_0(980)$ and $a_0(980)$ have long been 
identified as candidates for $K\bar{K}$ molecules. 
In the $b\bar{b}$ sector $B\bar{B}^*$, $B^*\bar{B}^*$  molecules \cite{tornqvist_x3872}
in addition to  a $B^*\bar{B_1}$ molecule bound state are expected
\cite{Cheung:2007wf,Lee:2007gs,Ding:2007ar}. 
In addition, threshold effects due to $\pi$-exchange and hybrid states
are also expected in both the $b\bar{b}$ and $s\bar{s}$ 
sectors \cite{Close:2008hv}. 
It is therefore of great interest
to see if there are corresponding exotic meson candidates
in the $b$-quark and $s$-quark sectors.  Some recent
results indicate that this may be the case.

\subsubsection{An anomalous partial width for 
``$\Upsilon(5S)$''$\rt\pi^+\pi^-\Upsilon(1S)$}

The bottomonium states are the $b\bar{b}$ counterparts of the charmonium 
mesons.  For these, the $J^{PC}=1^{--}$ states are the $\Upsilon(nS)$
mesons.  Most of the data accumulated by the KEKB and PEPII $B$-factory  
experiments is at the cm energy that corresponds to 
the peak of the $\Upsilon(4S)$, the $4^3S_1$ $b\bar{b}$ state
at 10,580~MeV/$c^2$, which is just above the threshold for producing $B\bar{B}$
meson pairs.  Using their large $\Upsilon(4S)$ data sample, the BaBar group measured 
the partial widths  for $\Upsilon(4S)\rt \pi^+\pi^- \Upsilon(2S)$
and $\pi^+\pi^-\Upsilon(1S)$ of $(1.8\pm 0.4)$~keV/$c^2$ and $(1.7\pm 
0.5)$~keV/$c^2$,  respectively~\cite{BaBar_4s2pipi1s}, the latter value
has been confirmed by Belle~\cite{Belle_4s2pipi1s} and both values are 
similar to those for dipion transitions from the $\Upsilon(3S)$ to the 
$\Upsilon(2S)$ ($0.6\pm0.2$~keV/$c^2$) and $\Upsilon(1S)$ 
($1.2\pm0.2$~keV/$c^2$)~\cite{PDG}.

\begin{figure}[htb!]
\centerline{\epsfxsize 4.0 truein
\epsfbox{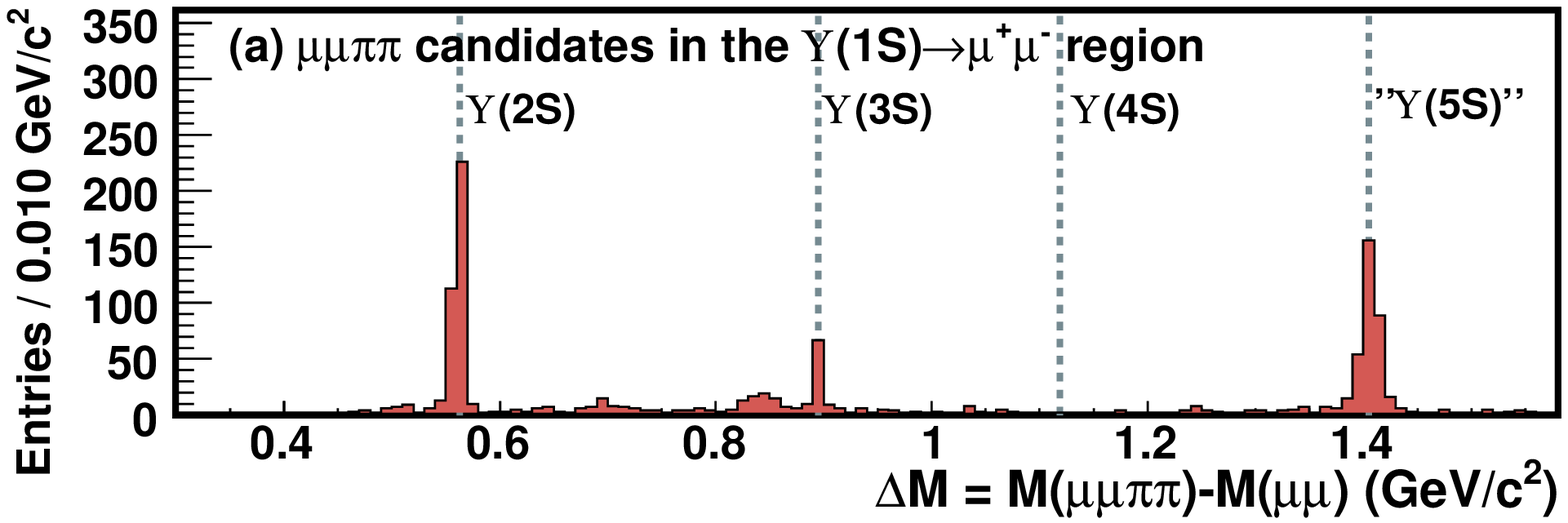 }}
   \vspace{1mm}
\centerline{\epsfxsize 4.0 truein
\epsfbox{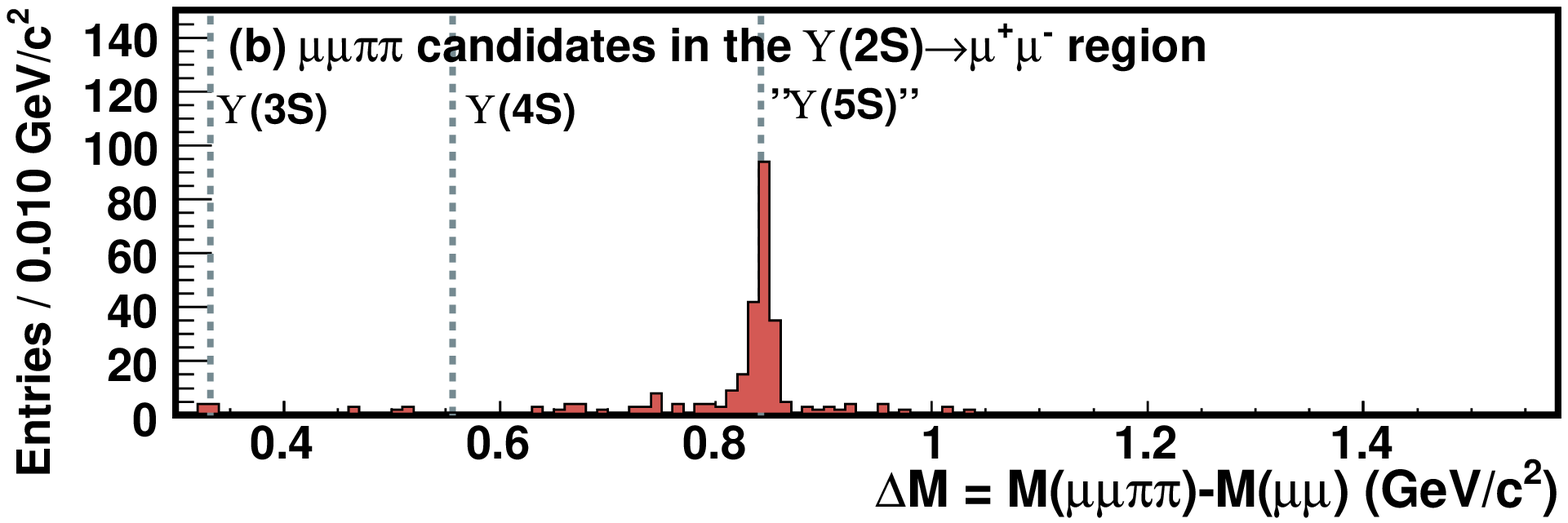 }}
\caption{\label{fig:belle_5s2pipi1s}
Belle's
$M(\mu^+\mu^-\pi^+\pi^-)-M(\mu^+\mu^-)$ mass difference distributions
for events with {\bf (a)} $M(\mu^+\mu^-)=\Upsilon(1S)$
and {\bf (b)} $M(\mu^+\mu^-)=\Upsilon(2S)$ (Ref.~\cite{Belle_5s2pipi1s}). 
Vertical dashed
lines show the expected locations for 
$\Upsilon(nS)\rt \pi^+\pi^-\Upsilon(1,2S)$ transitions.
(The $\Upsilon(2,3S)\rt\pi^+\pi^-\Upsilon(1S)$ signals 
in Fig.~\ref{fig:belle_5s2pipi1s}(a) are
produced by radiative-return transitions 
$e^+e^-\rt\gamma_{ISR}\Upsilon(2,3S)$.) }
\end{figure}

Recently, Belle accumulated a much smaller data sample at 10,870~MeV,
the peak of the $\Upsilon(5S)$, and found huge signals 
for $\pi^+\pi^-\Upsilon(1S)$, $\pi^+\pi^-\Upsilon(2S)$ and 
$\pi^+\pi^-\Upsilon(3S)$ (see Fig.~\ref{fig:belle_5s2pipi1s}).  
If these are attributed to
dipion transitions from the $\Upsilon(5S)$, the partial widths
are~\cite{Belle_5s2pipi1s}:
\begin{eqnarray}
\label{eq:pipi1s}
\Gamma(\hbox{``}\Upsilon(5S)\hbox{''}\rt\pi^+\pi^-\Upsilon(1S))&=& (590\pm 100)~{\rm 
keV}/c^2\\\nonumber
\Gamma(\hbox{``}\Upsilon(5S)\hbox{''}\rt\pi^+\pi^-\Upsilon(2S))&=& (850\pm 175)~{\rm 
keV}/c^2\\\nonumber
\Gamma(\hbox{``}\Upsilon(5S)\hbox{''}\rt\pi^+\pi^-\Upsilon(3S))&=& (520\pm 220)~{\rm 
keV}/c^2,
\end{eqnarray}
which are more than two-orders-of-magnitude larger than those for the 
corresponding transitions for the $\Upsilon(4S)$, $\Upsilon(3S)$
or $\Upsilon(2S)$.  A likely interpretation is that a $b\bar{b}$
counterpart of the $Y(4260)$, the $Y_b$,  may be overlapping the 
$\Upsilon(5S)$, and this is the source of the
anomalous $\pi^+\pi^-\Upsilon(nS)$ production~\cite{hou:2007}.\footnote{This 
is the reason
for the quotation marks around $\Upsilon(5S)$ in Eq.~\ref{eq:pipi1s}.}
As noted in Ref.~\cite{Belle_5s2pipi1s}, this hypothesis could be
verified by measuring the cm-energy dependence of the cross sections
for $e^+e^-\rt \pi^+\pi^-\Upsilon(nS)$ around 10,870~MeV.  
Another suggestion is that the anomalously high $\pi^-\pi^-$  
transitions could be due to the mixing of conventional $b\bar{b}$ states with 
thresholds and subsequent rescattering~\cite{Close:2008hv}.  This could
be tested by looking for other final states.

\subsubsection{The $Y(2175)\rt \phi f_0(980)$; an
$s\bar{s}$ counterpart of the $Y(4260)$?}

In a study of the ISR process 
$e^+e^-\rt\gamma_{ISR}\phi f_0(980)$, where $\phi\rt K^+K^-$ and 
$f_0(980)\rt\pi^+\pi^-$, the BaBar group observed a distinct
resonance-like peak in the $\phi f_0(980)$ invariant mass
distribution at a mass $M=(2175\pm18)$~MeV/$c^2$ with a
full width $\Gamma= (58\pm26)$~MeV/$c^2$~\cite{BaBar_y2175}. 
Recently, a $\phi f_0(980)$ invariant mass peak with mass and 
width values consistent with the BaBar observation was
seen in $J/\psi\rt \phi f_0(980)\eta$ decays by the 
BES group~\cite{BES_y2175}.  The 
$\phi$ meson is the $s\bar{s}$ counterpart of the $J/\psi$, and 
the observed structure, called the $Y(2175)$, has similar production
and decay characteristics as the $Y(4260)$.  This has led to some 
speculation 
that the $Y(2175)$ may be the $Y_s$, {\it i.e.} an $s$-quark system counterpart
to the $Y(4260)$~\cite{ding_y2175}.   While this is an intriguing
idea, the experimental situation is far from conclusive, and the
$Y(2175)$ may very well be an excited state of the $\phi$ or
some other $q\bar{q}$ meson.  
Ding and Yan \cite{Ding:2007pc} have suggested that studying the decay
modes of the $Y(2175)$ can distinguish between 
the conventional $2^3D_1 (s\bar{s})$ and strangeonium hybrid explanations.
Specifically, the dominant decay modes of an $s\bar{s}$ hybrid are
$K_1(1400)\bar{K}$ and $K_1(1270)\bar{K}$ with, for example, the decays 
to $K\bar{K}$ or $K^*\bar{K^*}$ 
forbidden.  
In contrast the $2^3D_1 (s\bar{s})$ is expected to have large 
branching fractions to $K\bar{K}$ 
and $K^*\bar{K^*}$.  The
$3^3S_1(s\bar{s})$ is not considered to be a candidate as it is predicted to be 
quite broad. 
Further studies of the properties
of the $Y(2175)$, and searches for $s$-quark counterparts of     
other $XYZ$ states could help clarify the situation.  

\subsection{Summary}

The $B$-factory experiments have  uncovered a large (and rapidly growing) number of 
candidates for charmonium and charmonium-like meson states, many of which cannot
be easily accommodated  by current theoretical expectations for $c\bar{c}$ mesons.   
A number of models have been proposed to explain these states, including 
meson-antimeson molecules, diquark-diantiquark bound states, $c\bar{c}$-gluon
hybrids and threshold effects.  None of the proposed mechanisms easily accounts
for all of the observations.  Moreover, there is some evidence for similar
behaviour in the $b$- and $s$-quark sectors.
  
As a summary, we list in Table~\ref{tab:xyz_states} the states 
discussed above together with some of their pertinent properties.

\begin{table}[htb!]
\caption{A summary of the properties of the candidate $XYZ$ mesons 
discussed in the text. For simplicity, the quoted errors are quadratic 
sums of statistical and systematic uncertainties.}
\label{tab:xyz_states}
\begin{tabular}{@{}lccccc@{}}
\toprule

\colrule
state    & $M$~(MeV) &$\Gamma$~(MeV)    & $J^{PC}$ & Decay Modes        & 
Production  Modes\\\hline
$Y_s(2175)$& $2175\pm8$&$ 58\pm26 $& $1^{--}$ & $\phi f_0(980)$     &  
$e^+e^-$~(ISR), 
$J/\psi$ decay \\
$X(3872)$& $3871.4\pm0.6$&$<2.3$& $1^{++}$ & $\pi^+\pi^- J/\psi$,$\gamma 
J/\psi$
 & $B\rt KX(3872)$, 
$p\bar{p}$ \\
$X(3875)$& $3875.5\pm 1.5$&$3.0^{+2.1}_{-1.7}$  &  &$D^0\bar{D^0}\pi^0$ & 
$B\rt 
K X(3875)$ \\
$Z(3940)$& $3929\pm5$&$ 29\pm10 $& $2^{++}$ & $D\bar{D}$          & 
$\gamma\gamma$        \\
$X(3940)$& $3942\pm9$&$ 37\pm17 $& $J^{P+}$ & $D\bar{D^*}$        & 
$e^+e^-\rt J
/\psi X(3940)$ \\
$Y(3940)$& $3943\pm17$&$ 87\pm34 $&$J^{P+}$ & $\omega J/\psi$     & $B\rt 
K Y(39
40)$          \\
$Y(4008)$& $4008^{+82}_{-49}$&$ 226^{+97}_{-80}$ &$1^{--}$& $\pi^+\pi^- 
J/\psi$ 
& $e^+e^-$(ISR)  \\
$X(4160)$& $4156\pm29$&$ 139^{+113}_{-65}$ &$J^{P+}$& $D^*\bar{D^*}$& 
$e^+e^-\rt
 J/\psi X(4160)$\\ 
$Y(4260)$& $4264\pm12$&$ 83\pm22$ &$1^{--}$&  $\pi^+\pi^- J/\psi$ & 
$e^+e^-$(ISR
)       \\
$Y(4350)$& $4361\pm13$&$ 74\pm18$ &$1^{--}$&  $\pi^+\pi^- \psi^{\prime}$ & 
$e^+e
^-$(ISR)       \\
$Z(4430)$& $4433\pm5$&$ 45^{+35}_{-18}$ & ? & $\pi^{\pm}\psi^{\prime}$ & 
$B\rt K
Z^{\pm}(4430)$\\
$Y(4660)$& $4664\pm12$&$ 48\pm15 $ &$1^{--}$&  $\pi^+\pi^- \psi^{\prime}$ 
& $e^+e^-$(ISR)       \\
$Y_b$     & $\sim 10,870$ & ?    &  $1^{--}$ &  $\pi^+\pi^-\Upsilon(nS)$
& $e^+e^-$       \\
\botrule
\end{tabular}
\end{table}

\section*{Summary points}

\begin{enumerate}
\item 
QCD-motivated quark potential models describe the properties of the
charmonium spectrum quite well.
\item 
In the last few years the $\eta_c'$, $h_c$ and 
$\chi^{\prime}_{c2}$ charmonium states 
have been discovered and their measured properties are in good agreement with
the quark model predictions.
\item 
There is accumulating evidence for the existence of mesons
with mass in the region between  3800~MeV/$c^2$ and 4700~MeV/$c^2$
that are not easily explained as simple quark-antiquark states
of the charmonium model. 
These mesons have a number of intriguing and/or unexpected properties.
\item
These states are relatively narrow although many of them are well above
relevant open-charm thresholds.   
Many of them have partial widths for decays to charmonium~+~light~hadrons
that are at the $\sim$MeV scale, which is much larger than is typical
for established $c\bar{c}$ charmonium meson states.
\item
The $X(3872)$, first seen as a narrow peak in the invariant mass distribution 
of $\pi^+\pi^- J/\psi$ in $B^-\to K^-\pi^+\pi^- J/\psi$, is not easily 
described as a conventional $c\bar{c}$ state and is a strong candidate
for a $D\bar{D^*}$ molecule.
\item
The new $1^{--}$ charmonium states are not apparent in the
$e^+e^-\rt$ charmed-meson-pair or the total hadronic cross sections
and there are no evident changes in the properties of these states 
at the $DD^{**}$ mass threshold.  
There seems to be some selectivity: states seen to decay to final
states with a $\psi^{\prime}$ are not seen in the corresponding $J/\psi$
channel, and {\it vice versa}.  At least one of these states is regarded
as a strong candidate for a charmonium hybrid.
\item
At least one of these new states, the $Z(4430)$,  is unique in that it has 
a non-zero electric charge. 
\item
There is some evidence that similar states exist in the $s$- and $b$-quark
sectors.
\end{enumerate}

\section*{Future issues}

\begin{enumerate}
\item  To confirm that these states are not conventional $c\bar{c}$ states,
more detailed studies of their properties, in particular, measurements of
their quantum numbers and measurements of other decay modes, are required.
\item It will also be important to put rigorous quantitative limits on the 
non-observation of final states.
\item The existence of similar states 
in the $s\bar{s}$ and $b\bar{b}$ sectors should be verified and their properties
measured.
\item Because many of these states have been observed to have masses 
that are close to kinematic 
thresholds it is necessary that we improve the theoretical understanding of
threshold effects, including $\pi$-exchange contributions, 
 coupled channel effects, and the interaction between both resonances
and thresholds via coupled channel effects and the resulting observed cross sections.

\end{enumerate}

\section*{Acknowledgements}
We thank T. Barnes, K.-F. Chen,
S.-K. Choi, F. Close, E. Swanson, K. Trabelsi,
S. Uehara and C.Z. Yuan for 
helpful communications. This work was supported in part the Natural 
Sciences and Engineering Research Council of Canada, the U.S. Department 
of Energy and the  Chinese Academy of Sciences.

\end{document}